\documentclass[9pt,twocolumn]{article}
\usepackage{times,bm}
\usepackage[affil-it,]{authblk}
\usepackage[pagebackref=false,colorlinks=true,linkcolor=blue,citecolor=red,urlcolor=acsblue]{hyperref}
\usepackage{geometry}
\usepackage[switch*, displaymath,pagewise, mathlines]{lineno}
\usepackage{graphicx}
\usepackage{cite}
\usepackage{amsmath}
\usepackage{amsfonts}
\usepackage{amssymb}
\usepackage{subfloat}
\usepackage{slashed}
\usepackage{color}
\usepackage{multirow,multicol}
\usepackage{tabulary}
\usepackage[flushleft]{threeparttable}
\usepackage{pgfplots,subfigure}
\usepackage{xcolor}
\numberwithin{equation}{section}
\usepackage{tikz}
\usepackage{ulem}
\usepackage{hyperref}
\usepackage{nameref}
\usepackage{comment}

\usepgflibrary{arrows}
\usetikzlibrary{shapes.callouts}
\tikzset{
	level/.style   = { thick, },
	connect/.style = { dotted, red   },
	notice/.style  = { draw, rectangle callout, callout relative pointer={#1} },
	label/.style   = { text width=2cm }
}

\usepackage{letltxmacro}
\makeatletter
\let\oldr@@t\r@@t
\def\r@@t#1#2{%
	\setbox0=\hbox{$\oldr@@t#1{#2\,}$}\dimen0=\ht0
	\advance\dimen0-0.2\ht0
	\setbox2=\hbox{\vrule height\ht0 depth -\dimen0}%
	{\box0\lower0.4pt\box2}}
\LetLtxMacro{\oldsqrt}{\sqrt}
\renewcommand*{\sqrt}[2][\ ]{\oldsqrt[#1]{#2}}
\makeatother
\usepackage{environ}

\definecolor{acsblue}{RGB}{17,76,139}

\begin{document}

\fontsize{7.6}{8.6}\selectfont

\newcommand{{\ri}}{{\rm{i}}}
\newcommand{{\Psibar}}{{\bar{\Psi}}}

\title{\mdseries{Geometry--gauge controlled dynamics and localization of a two-particle system on a helicoidal manifold}}
\author{ \textit {Abdullah Guvendi}$^{\ 1}$\footnote{\textit{E-mail: abdullah.guvendi@erzurum.edu.tr (Corresponding Author)} }~,~ \textit {Hassan Hassanabadi}$^{\ 2\,\ 3}$\footnote{\textit{E-mail: hha1349@gmail.com} }  \\
	\small \textit {$^{\ 1}$Department of Basic Sciences, Faculty of Science, Erzurum Technical University, 25050, Erzurum,
		Türkiye}\\
	\small \textit {$^{\ 2}$Physics Department, California State University, Fresno, CA 93740, USA}\\
	\small \textit {$^{\ 3}$Department of Physics, Faculty of Science, University of Hradec Králové, Rokitanského 62, 500 03 Hradec Králové, Czechia}}

\date{}
\maketitle

\begin{abstract}
{\fontsize{7.6}{8.6}\selectfont \setlength{\parindent}{0pt}
We study the dynamics and localization of two oppositely charged particles constrained to a helicoidal manifold in a uniform magnetic field. The embedding-induced metric and the pullback of the ambient electromagnetic gauge potential, together with restriction to the reflection-symmetric longitudinal rest frame, lead to an exact reduction of the relative dynamics to a one-dimensional Hamiltonian with a coordinate-dependent kinetic term and a gauge-shifted momentum. We analyze the corresponding effective potential, turning points, and classically allowed regions, and determine how the geometric and magnetic parameters modify the bounded relative motion. For a regularized attractive interaction, we derive the zero-energy condition for localization around the symmetric configuration and obtain the local stiffness governing its stability. When this stiffness changes sign while the quartic coefficient remains positive, the symmetric minimum undergoes a pitchfork-type bifurcation to two symmetry-related finite-separation minima of the relative coordinate. Canonical quantization of the reduced Hamiltonian then gives the low-energy spectrum in the harmonic approximation and the associated zero-energy localization thresholds. At the critical stiffness, the quadratic term vanishes and the quartic term provides the leading contribution to the local low-energy scaling. These results establish how helicoidal geometry and magnetic coupling modify the relative localization and low-energy spectral properties of the two-particle system.}
\end{abstract}

\begin{small}
\begin{center}
\textit{\fontsize{7.6}{8.6}\selectfont \setlength{\parindent}{0pt} Keywords: Two-body system; Helicoidal manifold; Constrained dynamics; Magnetic confinement; Localization }
\end{center}
\end{small}


\section{\mdseries{Introduction}}
\setlength{\parindent}{0pt}

Geometry has traditionally been regarded as a background structure in which physical processes occur, while dynamical control has been achieved primarily through external potentials, material parameters, or applied fields \cite{atanasov2015helicoidal,phong2022boundary,background,gurtas2025twist,zhang2014strain,guvendi2025damped,guvendi2025photonic,ferrari2008schrodinger}. Recent developments in curved quantum matter, nanostructured materials, photonic architectures, and synthetic quantum platforms have shown that the geometry of an underlying space can directly modify observable physical properties \cite{guvendi2023fermion,dogan2025geometric,dogan2025ray,gurtas2025ray,gurtas2025ray2}. In these systems, deformation is no longer only a constraint imposed on particles or waves; rather, it becomes a physical mechanism that can control transport, localization, topology, and spectral properties. The development of programmable curved surfaces and engineered gauge fields has therefore opened new possibilities for designing systems in which geometry directly influences dynamical responses. From the theoretical perspective, an important step toward this viewpoint was provided by the gauge-theoretical formulation of quantum motion in curved waveguides, where curvature, torsion, and transverse confinement were shown to generate geometry-dependent gauge structures and nonadiabatic mode couplings within a unified Hamiltonian framework \cite{stockhofe2014}. More recently, these ideas have been extended to quantum two-body motion on helicoidal geometries, demonstrating how helicoidal geometry modifies correlated quantum dynamics and wave-packet evolution \cite{gurtas2025twist,schmelcher2026}. Despite these advances, an important question remains: can geometric deformation itself act as a controllable mechanism for regulating the formation and organization of classical and quantum states without relying on externally imposed confinement mechanisms?

\vspace{0.03cm}

A central challenge in addressing this question is that curvature modifies dynamics in a fundamentally different manner from conventional scalar potentials. In flat systems, localization is typically achieved through externally introduced energy landscapes, whereas in curved spaces the metric directly modifies the kinetic structure, changes inertial response, and alters the accessible phase-space geometry \cite{guvendi2026charged}. When electromagnetic fields are introduced, this geometric modification of motion becomes further enriched because gauge potentials are projected onto the curved manifold and become coupled to the underlying metric structure. These developments have shown that the effective gauge structures generated by constrained motion are directly connected to the underlying geometry and transverse mode structure rather than being simple corrections to flat-space dynamics \cite{stockhofe2014}. Nevertheless, the combined influence of geometric gauge fields, external magnetic fields, and interparticle interactions on nonlinear localization and phase-transition phenomena remains largely unexplored. This produces a class of systems where curvature and gauge degrees of freedom cannot be treated as independent corrections but instead determine the resulting dynamical behavior. Although curvature effects have been explored in quantum particles, topological systems, and wave propagation \cite{lopes2012continuum,gonzalez2010graphene,errehymy2025frame,garcia2020graphene}, a systematic description of how geometry controls localization transitions, stability boundaries, and quantum spectral reconstruction remains incomplete \cite{guvendi2026phase}.

\vspace{0.03cm}

Building upon these geometric gauge formulations for curved quantum waveguides and helicoidal two-body systems \cite{gurtas2025twist,stockhofe2014,schmelcher2026}, we develop a curvature--gauge framework for a charged interacting two-body system subjected to an external magnetic field. In contrast to previous studies, our formulation incorporates the combined effects of Coulomb interaction, helicoidal geometry, and magnetic gauge coupling within a unified reduced Hamiltonian, allowing the nonlinear classical dynamics to be analyzed exactly and providing the corresponding quantum Hamiltonian governing the low-energy spectrum. The intrinsic helicoidal geometry provides an analytically tractable platform where the embedding-induced metric modifies the kinetic sector through a position-dependent dynamical response, while the projected gauge field generates a coordinate-dependent modification of the conserved momentum. By reducing the constrained dynamics to an exact effective Hamiltonian, we show that geometric deformation does not merely provide a perturbative correction to a flat-space system but reorganizes the phase-space structure. This curvature--gauge relation leads to several related phenomena, including geometry-induced reconstruction of classical trajectories, localization transitions at the zero-energy threshold, and symmetry-breaking quantum phase transitions associated with the renormalization of the effective restoring force. The helicoidal twist therefore emerges as a controllable geometric parameter capable of modifying the accessible phase space through changes in kinetic geometry and gauge coupling.

\vspace{0.03cm}

The quantum consequences of this geometric control mechanism extend beyond classical localization. The quantized effective Hamiltonian demonstrates that the same curvature-induced reconstruction responsible for classical phase-space transitions also governs the organization of bound states and excitation spectra. Previous studies showed that helicoidal geometry can strongly affect quantum two-body dynamics and wave-packet evolution \cite{schmelcher2026}. Here, we demonstrate that the inclusion of magnetic gauge coupling further modifies the low-energy spectrum and leads to geometry-dependent quantum critical behavior. In particular, the disappearance of the quadratic confinement scale produces a critical regime in which the low-energy spectrum changes from harmonic to quartic behavior, revealing a direct connection between geometric deformation and quantum critical dynamics (see also \cite{guvendi2026phase}). Our results show that engineered curvature together with gauge structure provides a route for controlling localization, stability, and spectral properties in quantum systems. Beyond the helicoidal realization considered here, this framework applies to curved electronic materials, photonic structures, metamaterials, and synthetic quantum platforms where geometry and gauge fields can be designed to control physical states.

\vspace{0.03cm}

The organization of this paper is as follows. In Sec.~\ref{TBD}, we develop the exact classical formulation of the helicoidal two-particle system by revisiting the embedding-induced metric, the projected electromagnetic gauge structure, and the corresponding reduced Hamiltonian dynamics. We analyze the curvature--gauge modification of the phase-space structure and identify the resulting nonlinear classical regimes through the effective potential landscape and turning-point analysis. In Sec.~\ref{ZET}, we investigate the zero-energy localization mechanism generated by the competition between attractive interaction, geometric deformation, and magnetic gauge coupling. The emergence of localized trajectories, the geometry--gauge controlled stiffness renormalization, and the associated bifurcation structure are examined through analytical expansions and phase-space characterization. In Sec.~\ref{PT}, we study the curvature--gauge controlled phase transitions and dynamical signatures by introducing suitable dimensionless control parameters and determining the critical boundaries separating distinct localization regimes. In Sec.~\ref{QS}, we extend the analysis to the quantum domain by quantizing the reduced Hamiltonian and demonstrating the geometry-induced reconstruction of the low-energy spectrum, including the critical transition from harmonic to quartic confinement and the geometry-dependent localization thresholds of quantum states. Finally, Sec.~\ref{conc} summarizes the main findings and discusses the broader implications of curvature and gauge engineering for controlling classical and quantum states in curved spaces.

\section{\mdseries{Two-Particle Dynamics}}\label{TBD}

\setlength{\parindent}{0pt}

We consider a constrained two-body system of identical classical point particles of mass $\mu$, evolving on a helicoidal Riemannian submanifold $\mathcal{M}\subset\mathbb{R}^3$, where curvature, topology, and gauge structure are dynamically intertwined through the embedding-induced metric and electromagnetic pullback. The helicoidal surface possesses a strictly negative but spatially varying Gaussian curvature (for non-zero twist density) which enforces a non-Euclidean kinetic structure and prohibits global inertial factorization of the dynamics. The particles carry opposite charges \(q_1=+q,\, q_2=-q\) and this removes the uniform center-of-mass contribution in the symmetric relative sector. The system is placed in a uniform magnetic field \cite{guvendi2026phase}
\begin{equation}
\vec{\mathcal B}=\mathcal B\,\hat{x}\label{UMF}
\end{equation}
whose pullback onto $\mathcal{M}$ through the embedding map $\iota:\mathcal{M}\hookrightarrow\mathbb{R}^{3}$ produces a coordinate-dependent intrinsic gauge structure and an effective surface magnetic field that varies with the helicoidal coordinate. The helicoidal embedding is defined as \cite{background,gurtas2025twist}
\begin{equation}
\vec{r}(\xi,v)=\big(v,\;\xi\cos(\omega v),\;\xi\sin(\omega v)\big)
\end{equation}
with winding density \(\omega=\frac{2\pi m}{L}\), which controls the twist rate of the helicoidal embedding along the longitudinal direction. The intrinsic curvature of the helicoidal manifold follows directly from the metric tensor
\begin{equation}
g_{ij}=\textrm{diag}\left(1,\chi(\xi)\right),\qquad \chi(\xi)=1+\omega^2\xi^2 .
\end{equation}
For a two-dimensional metric of the form
\begin{equation}
ds^2=d\xi^2+\chi(\xi)\,dv^2 ,\label{eq:metric}
\end{equation}
the Gaussian curvature is given by
\begin{equation}
K(\xi)=-\frac{(\sqrt{\chi})''}{\sqrt{\chi}}\Rightarrow K(\xi)
=-\frac{\omega^2}{\left(1+\omega^2\xi^2\right)^2}.
\end{equation}
Therefore, for any nonzero winding density,
\begin{equation}
\omega\neq0,\qquad K(\xi)<0 ,
\end{equation}
showing that the helicoidal embedding generates an intrinsically negatively curved geometry with spatially varying Gaussian curvature. The curvature is spatially varying rather than constant, with its magnitude decreasing away from the helicoidal axis as
\begin{equation}
|K(\xi)|\sim \xi^{-4},\qquad |\xi|\rightarrow\infty .
\end{equation}
Hence, the twist-induced geometry produces a spatially varying negative Gaussian curvature distribution that modifies the intrinsic kinetic structure of the constrained dynamics. The electromagnetic potential in symmetric gauge \cite{guvendi2026phase}
\begin{equation}
\vec{\mathcal A}=\frac{\mathcal B}{2}(0,-z,y)
\end{equation}
satisfies
\begin{equation}
\nabla\times\vec{\mathcal A}=\mathcal B\hat{x}
\end{equation}
and reduces on $\mathcal{M}$ to the intrinsic form
\begin{equation}
\mathcal A_\xi=0,\qquad \mathcal A_v(\xi)=\frac{\mathcal B\omega}{2}\xi^2.\label{GP}
\end{equation}
The quadratic dependence originates from the pullback of the ambient symmetric-gauge potential onto the helicoidal embedding. The full gauge-covariant Lagrangian is
\begin{equation}
\mathcal L=
\sum_{i=1}^2\left[
\frac{\mu}{2}\left(\dot\xi_i^2+\chi(\xi_i)\dot v_i^2\right)
+q_i\mathcal A_v(\xi_i)\dot v_i
\right]
-\frac{k}{2}(\xi_1-\xi_2)^2
\end{equation}
where the interaction is chosen harmonic to isolate geometric and gauge nonlinearities. Substituting $q_1=+q$, $q_2=-q$ yields
\begin{equation}
\begin{split}
\mathcal L&=\frac{\mu}{2}\left(\dot\xi_1^2+\dot\xi_2^2+\chi(\xi_1)\dot v_1^2+\chi(\xi_2)\dot v_2^2\right)\\
&+q\left[\mathcal A_v(\xi_1)\dot v_1-\mathcal A_v(\xi_2)\dot v_2\right]-\frac{k}{2}(\xi_1-\xi_2)^2  
\end{split}
\end{equation}
which explicitly exhibits the opposite-sign minimal coupling of the two charged particles under the exchange $1\leftrightarrow2$. Introducing center-of-mass and relative coordinates
\begin{equation}
\Xi=\frac{\xi_1+\xi_2}{2},\quad \xi=\xi_1-\xi_2,\quad V=\frac{v_1+v_2}{2},\quad v=v_1-v_2
\end{equation}
with inverse relations
\begin{equation}
\xi_1=\Xi+\frac{\xi}{2},\quad \xi_2=\Xi-\frac{\xi}{2},\quad v_1=V+\frac{v}{2},\quad v_2=V-\frac{v}{2}
\end{equation}
the transverse kinetic sector decomposes exactly as
\begin{equation}
\dot\xi_1^2+\dot\xi_2^2=2\dot\Xi^2+\frac{1}{2}\dot\xi^2
\end{equation}
while the longitudinal sector retains full nonlinear curvature coupling
\begin{equation}
\begin{split}
 &\chi(\xi_1)\dot v_1^2+\chi(\xi_2)\dot v_2^2=\\
&\chi\!\left(\Xi+\frac{\xi}{2}\right)\left(\dot V+\frac{\dot v}{2}\right)^2+
\chi\!\left(\Xi-\frac{\xi}{2}\right)\left(\dot V-\frac{\dot v}{2}\right)^2   
\end{split}
\end{equation}
and the electromagnetic sector becomes
\begin{equation}
q\left[\mathcal A_v\!\left(\Xi+\frac{\xi}{2}\right)\left(\dot V+\frac{\dot v}{2}\right)-
\mathcal A_v\!\left(\Xi-\frac{\xi}{2}\right)\left(\dot V-\frac{\dot v}{2}\right)\right]
\end{equation}
demonstrating that the coordinate-dependent metric and gauge potential couple the collective and relative coordinates, preventing their complete separation in the general sector. Because the Lagrangian is independent of $v_i$, each longitudinal coordinate is cyclic and the corresponding canonical momenta are conserved,
\[p_{v_i}=\mu\chi(\xi_i)\dot v_i+q_i\mathcal A_v(\xi_i),\qquad\dot p_{v_i}=0 ,\]
despite the coordinate dependence induced by the helicoidal geometry. Restricting to the reflection-symmetric submanifold $\Xi=0$ and imposing the invariant condition $\dot\Xi=0$ gives
\(\xi_1=\xi/2,\,\xi_2=-\xi/2\), and the reduced Lagrangian becomes
\begin{equation}
\mathcal L_{\mathrm{rel}}=
\frac{\mu}{4}\dot\xi^2+
\frac{\mu}{4}\chi\!\left(\frac{\xi}{2}\right)\dot v^2+
q\,\mathcal A_v\!\left(\frac{\xi}{2}\right)\dot v-\frac{k}{2}\xi^2
\end{equation}
where the gauge term survives because the opposite charges eliminate the center-of-mass gauge contribution while retaining the relative gauge coupling. The conjugate momentum is therefore
\begin{equation}
p_v=\frac{\mu}{2}\chi\!\left(\frac{\xi}{2}\right)\dot v+q\,\mathcal A_v\!\left(\frac{\xi}{2}\right)
\end{equation}
and inversion yields
\begin{equation}
\dot v=\frac{2}{\mu\,\chi(\xi/2)}\left[p_v-q\,\mathcal A_v\!\left(\frac{\xi}{2}\right)\right]
\end{equation}
which explicitly shows curvature-modulated gauge coupling (see also \cite{kolovs2023charged,narzilloev2021dynamics,aliev2002motion,turimov2022circular}). The Hamiltonian obtained from the Legendre transformation preserves the canonical symplectic structure, while the helicoidal geometry and gauge coupling appear through a coordinate-dependent effective kinetic term in phase space. No approximation in the twist parameter or magnetic coupling has been introduced; therefore the curvature and gauge contributions are retained exactly within the reduced classical model. Starting from the exact Legendre transform of the reduced Lagrangian, the Hamiltonian reads \cite{goldstein1950classical}
\begin{figure*}[ht]
\centering
\includegraphics[scale=0.50]{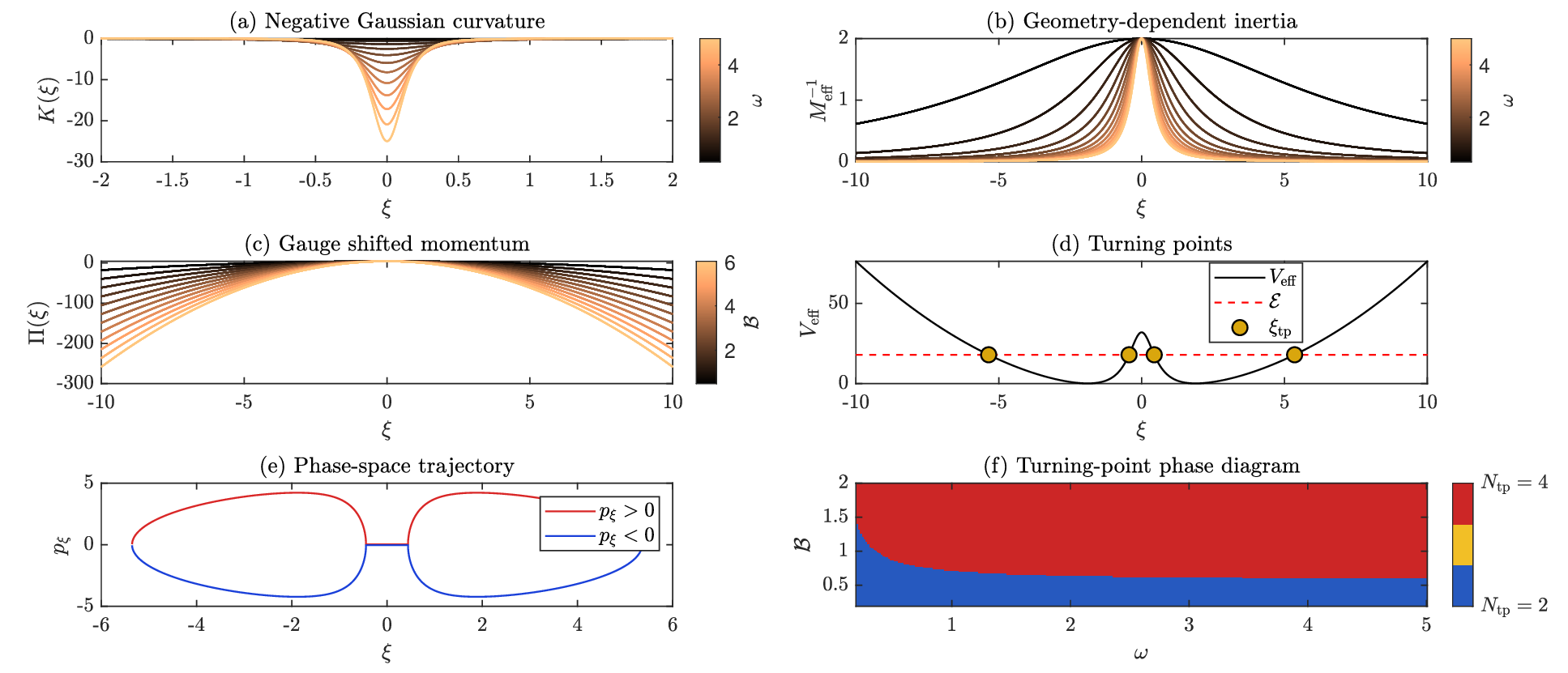}\\
\caption{\fontsize{7.6}{8.6}\selectfont   \textbf{Geometric and dynamical signatures of the helicoidal two-particle system}. The results are obtained from the exact reduced Hamiltonian
\(\mathcal H=\frac{p_\xi^2}{\mu}+\kappa\xi^2+\frac{(p_v-\alpha\xi^2)^2}{\mu(1+\Omega^2\xi^2)}\), with $\mu=q=1$, $k=0.08$ ($\kappa=0.04$), $p_v=4$, $\mathcal E=18$, and the reference parameters $\omega=3.5$ and $\mathcal B=2.5$ ($\Omega=1.75$, $\alpha=1.09375$). The color scales in panels (a)--(c) indicate the corresponding values of the control parameters $\omega$ and $\mathcal B$. (a) Negative Gaussian curvature $K(\xi)=-\omega^2/(1+\omega^2\xi^2)^2$ for different twist densities $\omega$. (b) Geometry-induced inverse effective inertia $M_{\rm eff}^{-1}=2/[\mu(1+\Omega^2\xi^2)]$ for different twist densities $\omega$. (c) Gauge-shifted canonical momentum $\Pi(\xi)=p_v-\alpha\xi^2$
for different magnetic fields $\mathcal B$. (d) Curvature-magnetic effective potential $V_{\rm eff}(\xi)$ together with the four physical turning points
obtained from the exact condition $\Phi(\xi)=\mathcal E-V_{\rm eff}(\xi)=0$. (e) Corresponding phase portrait $p_\xi=\pm\sqrt{\mu(\mathcal E-V_{\rm eff})}$,
where red and blue branches denote positive and negative momenta, respectively. (f) Physical turning-point phase diagram obtained by numerically counting the zeros of $\Phi(\xi)$ in the original coordinate $\xi$, revealing the regions with $N_{\rm tp}=2$, and $4$ turning points in the $(\omega,\mathcal B)$ parameter space.}
\label{fig:1}
\end{figure*}
\begin{equation}
\mathcal H=
\frac{p_\xi^2}{\mu}
+
\frac{\left[p_v-q\,\mathcal A_v\!\left(\frac{\xi}{2}\right)\right]^2}{\mu\,\chi\!\left(\frac{\xi}{2}\right)}
+
\frac{k}{2}\xi^2,\label{eq:hamiltonian}
\end{equation}
where the geometric factor and gauge projection are defined by
\begin{equation}
\chi\!\left(\frac{\xi}{2}\right)=1+\frac{\omega^2\xi^2}{4},\qquad
\mathcal A_v\!\left(\frac{\xi}{2}\right)=\frac{\mathcal B\omega}{8}\xi^2.
\end{equation}
The Hamiltonian is thus of the general form
\begin{equation}
\mathcal H = \mathcal T_\xi + \mathcal T_v + \mathcal V_{\mathrm{harm}},
\end{equation}
where the longitudinal kinetic sector is not quadratic in velocity alone but instead inherits a momentum-shifted, curvature-weighted structure that cannot generally be reduced to a flat kinetic form while simultaneously eliminating the curvature-induced gauge structure. The Hamiltonian flow is generated by the exact symplectic structure on the reduced phase space $(\xi,p_\xi)$, while the longitudinal Killing sector contributes only through the conserved parameter $p_v$, which acts as a frozen control variable rather than a dynamical degree of freedom. The canonical equations are therefore
\begin{equation}
\dot{\xi}=\frac{\partial \mathcal H}{\partial p_\xi}=\frac{2p_\xi}{\mu},\qquad
\dot{p}_\xi=-\frac{\partial \mathcal H}{\partial \xi},
\end{equation}
where the second equation must be interpreted as a geometrically dressed force balance rather than a simple Newtonian law. To expose the full nonlinear structure in a controlled way, it is useful to decompose the Hamiltonian force into three irreducible contributions: a harmonic restoring sector, a curvature-renormalization sector, and a gauge-coupling sector. Writing this decomposition explicitly,
\begin{equation}
\dot{p}_\xi=-k\xi-\frac{\partial}{\partial \xi}
\left[\frac{\left(p_v-q\,\mathcal A_v(\xi/2)\right)^2}{\mu\,\chi(\xi/2)}\right],
\end{equation}
which separates the radial dynamics into a harmonic restoring contribution and a combined curvature--gauge coupling term. In the reduced sector, the magnetic field does not appear as an explicit Lorentz-force term; instead, it modifies the dynamics through the gauge-induced shift of the conserved canonical momentum, \(p_v\rightarrow p_v-q\,\mathcal A_v(\xi/2)\), while the helicoidal geometry enters through the coordinate-dependent inertial factor $\chi(\xi/2)$. Consequently, the longitudinal sector is characterized by the effective inverse inertial coefficient \cite{guvendi2026phase}
\begin{equation}
M_{\mathrm{eff}}^{-1}(\xi)\equiv\frac{2}{\mu}\frac{1}{\chi(\xi/2)},
\end{equation}
which is modulated by the coordinate-dependent metric coefficient induced by the helicoidal embedding and coupled to the magnetic field through the shifted canonical momentum. For analytic transparency and structural compression, we introduce the geometric-magnetic control scales
\begin{equation}
\Omega \equiv \frac{\omega}{2},\qquad \alpha \equiv \frac{q\mathcal B\omega}{8},\qquad
\kappa \equiv \frac{k}{2},\qquad \Pi(\xi)\equiv p_v-\alpha \xi^2,
\end{equation}
so that the Hamiltonian assumes the compact non-Euclidean form
\begin{equation}
\mathcal H=\frac{p_\xi^2}{\mu}+\kappa \xi^2+\frac{1}{\mu}\frac{\Pi(\xi)^2}{1+\Omega^2\xi^2}.
\end{equation}
In this representation, the canonical force equation becomes a single rational nonlinear functional derivative,
\begin{equation}
\dot{p}_\xi=-2\kappa \xi-\frac{1}{\mu}\frac{d}{d\xi}\left(\frac{\Pi(\xi)^2}{1+\Omega^2\xi^2}\right),
\end{equation}
where all curvature and gauge effects are encoded in the composite functional $\Pi(\xi)$ and the metric factor $(1+\Omega^2\xi^2)^{-1}$. The system admits a strict first integral $\mathcal H=\mathcal E$, yielding the exact reduction within the symmetric invariant sector,
\begin{equation}
\dot{\xi}^2=
\frac{4}{\mu}\left[\mathcal E-\kappa\xi^2-\frac{1}{\mu}\frac{\Pi(\xi)^2}{1+\Omega^2\xi^2}\right],
\end{equation}
so that the entire phase portrait is encoded in the zero set of the curvature-dressed energy defect function
\begin{equation}
\Phi(\xi)\equiv
\mathcal E-\kappa\xi^2-\frac{1}{\mu}\frac{(p_v-\alpha\xi^2)^2}{1+\Omega^2\xi^2}.
\end{equation}
Turning points are therefore defined by $\Phi(\xi)=0$, which becomes algebraically exact upon clearing the geometric denominator:
\begin{equation}
\left(\mathcal E-\kappa\xi^2\right)\left(1+\Omega^2\xi^2\right)=
\frac{1}{\mu}(p_v-\alpha\xi^2)^2.
\end{equation}
Introducing the invariant quadratic coordinate \(x\equiv \xi^2\), the turning-point condition reduces exactly to a biquadratic spectral equation
\begin{equation}
\mathcal A_2 x^2+\mathcal A_1 x+\mathcal A_0=0,
\end{equation}
with curvature-gauge-energy dressed coefficients
\begin{equation}
\mathcal A_2=
-\kappa\Omega^2-\frac{\alpha^2}{\mu},\quad \mathcal A_1=
\mathcal E\Omega^2-\kappa+\frac{2\alpha p_v}{\mu},
\end{equation}
\begin{equation}
\mathcal A_0=
\mathcal E-\frac{p_v^2}{\mu}.
\end{equation}
The discriminant \(\Delta=\mathcal A_1^2-4\mathcal A_2\mathcal A_0\) determines the number of real roots of the algebraic turning-point equation in the invariant coordinate $x$. The resulting dynamical regimes are classified according to the nature of the admissible roots of this equation. For $\Delta<0$, the turning-point equation has no real solutions in the reduced coordinate $x$, implying that no finite turning points exist within the reduced coordinate description. At $\Delta=0$, two turning points merge, producing a critical configuration associated with a separatrix and the onset of marginal dynamical stability. For $\Delta>0$, real turning-point candidates emerge, and whenever the corresponding roots satisfy $x\geq0$, the motion is confined within a finite region of the helicoidal coordinate by the effective potential landscape. In the confining regime $k>0$, the quadratic coefficient
\begin{equation}
\mathcal A_2=-\kappa\Omega^2-\frac{\alpha^2}{\mu}
\end{equation}
is strictly negative, revealing the asymptotic dominance of the curvature and magnetic contributions in the turning-point structure. These contributions modify the asymptotic behavior of the effective potential and influence the large-amplitude accessibility of the radial coordinate. Nevertheless, the mixed coupling term \(4\alpha p_v/\mu\) contained in $\mathcal A_1$ introduces a directional bias controlled by the relative orientation between the conserved longitudinal canonical momentum and the magnetic flux. This term can continuously reshape the effective potential, either reinforcing or weakening the confinement depending on the sign of the
gauge--momentum coupling. Within this framework, the helicoidal embedding is not treated as a perturbative geometrical correction but as an intrinsic generator of the nonlinear phase-space structure. The induced metric factor $\chi(\xi)$ and the projected gauge potential $\mathcal A_v(\xi)$ determine a nonlinear Hamiltonian flow characterized by a coordinate-dependent kinetic term and a momentum-dependent effective potential landscape.

\setlength{\parindent}{0pt}

Figure~\ref{fig:1} demonstrates that the helicoidal geometry and the electromagnetic gauge structure generate a genuinely non-Euclidean Hamiltonian dynamics rather than acting as perturbative corrections to a flat-space two-particle system. The first important feature is the intrinsic negative curvature distribution, which remains strictly negative for any nonzero twist density and is spatially concentrated around the helicoidal axis. Consequently, the twist parameter $\omega$ controls both the magnitude of the geometric deformation and the spatial scale over which the effective inertial properties of the particles are modified. This effect is explicitly reflected in the coordinate-dependent inverse effective inertia, demonstrating that the helicoidal metric continuously redistributes the kinetic response along the relative coordinate. Therefore, the particle motion evolves on a position-dependent inertial background even before the electromagnetic contribution is introduced. The gauge sector generates an additional nonlinear deformation through the conserved canonical momentum shift. Unlike the conventional flat-space description, where the magnetic field appears through an explicit Lorentz-force term, the reduced dynamics incorporates the electromagnetic contribution through the curvature-dependent canonical displacement \(p_v\rightarrow p_v-\alpha\xi^2\), while the helicoidal geometry enters through the metric factor \((1+\Omega^2\xi^2)^{-1}\). The resulting effective potential simultaneously contains the harmonic restoring contribution, the geometry-induced inertial renormalization, and the gauge-induced momentum displacement. These contributions cannot be independently separated without eliminating the intrinsic information encoded by the helicoidal embedding. Hence, the accessible phase-space structure is governed by a nonlinear coupling between curvature, confinement, and magnetic gauge effects. A central result of Fig.~\ref{fig:1} is the emergence of a four-turning-point dynamical regime. The turning points are obtained from the condition \(\Phi(\xi)=\mathcal E-V_{\rm eff}(\xi)=0\), and therefore represent genuine boundaries of the classically allowed motion. The appearance of four physical roots, \(\xi_1<\xi_2<0<\xi_3<\xi_4\), reveals that the curvature-magnetic coupling generates a nontrivial effective landscape containing multiple bounded regions of motion. In phase space, this structure produces two momentum branches, \(p_\xi=\pm\sqrt{\mu(\mathcal E-V_{\rm eff})}\), where the positive and negative momentum trajectories describe the two orientations of motion on the same energy manifold. Thus, the accessible phase-space structure is not determined solely by the harmonic interaction but emerges from the nonlinear feedback between the helicoidal metric and the projected gauge field. The \((\omega,\mathcal B)\) phase diagram provides a global characterization of this dynamical reorganization. The transitions between the regions with \(N_{\rm tp}=2\), and \(N_{\rm tp}=4\) turning points represent changes in the structure of the accessible coordinate domain controlled by the geometric and electromagnetic parameters. Increasing the twist density modifies the curvature distribution and simultaneously strengthens the gauge coupling through \(\alpha=q\mathcal B\omega/8\), while increasing the magnetic field directly enhances the momentum displacement term \(\alpha\xi^2\). Their combined action transforms the effective potential landscape and can produce additional bounded motion intervals separated by dynamical barriers. This demonstrates that helicoidal geometry acts as an intrinsic generator of nonlinear localization: the spatial curvature modifies the kinetic structure, the gauge projection reshapes the canonical momentum, and their coupling produces dynamical regimes that are absent in planar systems. The helicoidal two-particle model therefore provides an exactly tractable framework for geometry-induced nonlinear phase-space engineering, where curvature, gauge structure, and classical Hamiltonian dynamics become intrinsically coupled.
\begin{figure*}[ht]
\centering
\includegraphics[scale=0.50]{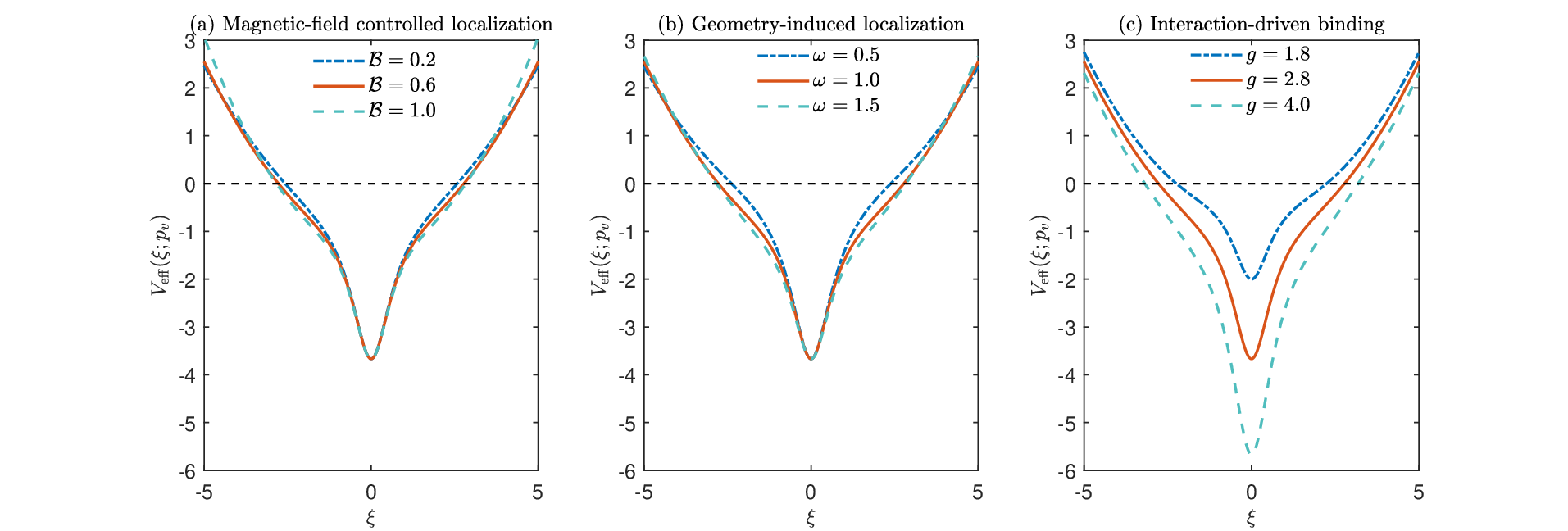}\\
\caption{\fontsize{7.6}{8.6}\selectfont   \textbf{Geometry--gauge controlled zero-energy localization.} The effective potential governing the zero-energy trajectories is shown for 
$p_v=1$, $\mu=1$, $\kappa=0.12$, $q=1$, and $a=0.6$.  The black dashed line denotes the zero-energy boundary $V_{\mathrm{eff}}(\xi;p_v)=0$, while the regions with 
$V_{\mathrm{eff}}(\xi;p_v)<0$ correspond to the classically accessible localization domains.  (a) Magnetic-field dependence for $\omega=1$ and $g=2.8$ with $\mathcal B=0.2$, $0.6$, and $1.0$.  (b) Twist-density dependence for $\mathcal B=0.6$ and $g=2.8$ with $\omega=0.5$, $1.0$, and $1.5$. (c) Interaction-strength dependence for $\mathcal B=0.6$ and $\omega=1$ with $g=1.8$, $2.8$, and $4.0$.  Here $\alpha=q\mathcal B\omega/8$ and $\Omega=\omega/2$.}
\label{fig:2}
\end{figure*}

\section{\mdseries{Geometry--Gauge Controlled Localization at the Zero-Energy Threshold}} \label{ZET}

\setlength{\parindent}{0pt}

We now investigate the classical zero-energy manifold $\mathcal H=0$ of the reduced two-particle system. The particles possess identical mass $\mu$ and opposite charges $q_1=+q$ and $q_2=-q$, and evolve on the helicoidal Riemannian manifold. The uniform magnetic field in equation \eqref{UMF} induces, after pullback onto $\mathcal M$, the effective gauge potential in equation \eqref{GP} which couples the electromagnetic response directly to the geometric deformation of the surface. Replacing the harmonic interaction by a finite-range attractive interaction by assuming that the interaction depends only on the relative transverse separation, the relative dynamics is governed by \cite{Downing,Guvendi}
\begin{equation}
\mathcal V_{\mathrm{int}}(\xi)=-\frac{g}{\sqrt{\xi^{2}+a^{2}}},\qquad g>0, \label{eq:SRI}
\end{equation}
where the parameter $a$ regularizes the short-distance behavior and preserves smoothness of the effective potential at the helicoidal axis. After elimination of the coordinate $v$, the exact reduced Hamiltonian becomes
\begin{equation}
\mathcal H=\frac{p_{\xi}^{2}}{\mu}+\kappa\xi^{2}+\frac{1}{\mu}\frac{(p_v-\alpha\xi^{2})^{2}}{1+\Omega^{2}\xi^{2}}-\frac{g}{\sqrt{\xi^{2}+a^{2}}},
\end{equation}
where
\begin{equation}
\kappa=\frac{k}{2},\qquad\Omega=\frac{\omega}{2},\qquad
\alpha=\frac{q\mathcal B\omega}{8},
\end{equation}
and $p_v$ is the conserved momentum associated with the Killing direction of the helicoidal geometry. The zero-energy manifold is defined by (see also \cite{Downing,Guvendi})
\begin{equation}
\mathcal H=0,
\end{equation}
which gives the exact radial constraint
\begin{equation}
p_{\xi}^{2}=-\mu V_{\mathrm{eff}}(\xi;p_v),
\end{equation}
with effective potential
\begin{equation}
V_{\mathrm{eff}}(\xi;p_v)=\kappa\xi^{2}+\frac{1}{\mu}\frac{(p_v-\alpha\xi^{2})^{2}}{1+\Omega^{2}\xi^{2}}-\frac{g}{\sqrt{\xi^{2}+a^{2}}}.
\end{equation}
Consequently, the classically accessible region is determined by
\begin{equation}
V_{\mathrm{eff}}(\xi;p_v)\leq0.
\end{equation}
In Figure \ref{fig:2}, the emergence of zero-energy localized trajectories is governed by the competition between the attractive interaction and the geometry--gauge modified effective kinetic contribution. The parameter $g$ sets the strength of the attractive energy scale and determines whether the effective potential can develop a classically accessible region with $V_{\mathrm{eff}}<0$, while the magnetic field and twist density modify the localization landscape through the coupled quantities $\alpha=q\mathcal B\omega/8$ and $\Omega=\omega/2$. Variations of the magnetic field and helicoidal deformation reshape the effective momentum barrier through the conserved longitudinal sector, thereby controlling the spatial extent and depth of the zero-energy localization region. Consequently, the magnetized helicoidal geometry provides an intrinsic mechanism for localization control by continuously modifying the zero-energy manifold rather than introducing an external confining potential. The existence of a zero-energy localized configuration therefore requires the attractive interaction to overcome the positive contributions arising from the geometry-dependent kinetic term and the harmonic confinement. To characterize the local structure of the binding region, we expand the effective potential around the helicoidal axis,
\begin{equation}
|\xi|\ll \min(\omega^{-1},a).
\end{equation}
obtaining the normal form
\begin{equation}
V_{\mathrm{eff}}(\xi;p_v)=C_0+A\xi^{2}+B\xi^{4}+\mathcal O(\xi^{6}),
\end{equation}
where
\begin{equation}
C_0=\frac{p_v^{2}}{\mu}-\frac{g}{a},
\end{equation}
\begin{equation}
A=\kappa-\frac{1}{\mu}\left(2\alpha p_v+\Omega^{2}p_v^{2}\right)+\frac{g}{2a^{3}},
\end{equation}
and
\begin{equation}
B=\frac{1}{\mu}\left(\alpha^{2}+2\alpha p_v\Omega^{2}+\Omega^{4}p_v^{2}\right)-\frac{3g}{8a^{5}}.
\end{equation}
These coefficients provide a direct separation of the competing physical mechanisms. $A$ represents the local stiffness renormalized by the helicoidal metric, the magnetic momentum shift, confinement,and attraction. A necessary condition for zero-energy localization is obtained from the value of the effective potential at the origin,
\begin{equation}
V_{\mathrm{eff}}(0)= C_0<0,
\end{equation}
which gives
\begin{equation}
\frac{p_v^{2}}{\mu}<\frac{g}{a}.
\end{equation}
This relation defines the longitudinal momentum threshold for the existence of zero-energy bound trajectories. The result demonstrates that curvature and magnetic coupling do not generate binding independently; rather, they redistribute the kinetic energy through the geometry-dependent longitudinal sector, while the attractive interaction provides the negative energy scale required for localization. The stationary configurations satisfy
\begin{equation}
\frac{\partial V_{\mathrm{eff}}}{\partial\xi}=0,
\end{equation}
which in the quartic approximation becomes
\begin{equation}
2A\xi+4B\xi^{3}=0.
\end{equation}
Hence the equilibrium points are
\begin{equation}
\xi=0,
\end{equation}
and, when allowed,
\begin{equation}
\xi_{\star}^{2}=-\frac{A}{2B}.
\end{equation}
For $A>0$ and $B>0$, the helicoidal axis remains a stable localization point. Conversely, for $A<0$ and $B>0$, the axial configuration becomes unstable and two symmetry-related minima appear at $\pm\xi_{\star}$. This bifurcation reflects a geometry-controlled restructuring of the effective potential landscape while preserving the reflection symmetry $\xi\rightarrow-\xi$ of the underlying helicoidal embedding. Near the stable minimum at $\xi=0$, the effective potential admits the harmonic
expansion
\begin{equation}
V_{\mathrm{eff}}(\xi;p_v)
\simeq
C_0+A\xi^2 ,
\end{equation}
where $A>0$ ensures local stability of the helicoidal axis. Due to the
non-standard kinetic normalization of the reduced Hamiltonian,
\begin{equation}
\mathcal H_\xi=\frac{p_\xi^2}{\mu}+A\xi^2+C_0 ,
\end{equation}
the local radial dynamics corresponds to an effective oscillator with frequency
\begin{equation}
\omega_{\mathrm{eff}}=2\sqrt{\frac{A}{\mu}} .
\end{equation}
The classical turning points of the zero-energy trajectories are determined by
\begin{equation}
V_{\mathrm{eff}}(\xi;p_v)=0 ,
\end{equation}
which, within the harmonic approximation, gives
\begin{equation}
\xi_{\max}^{2}=-\frac{C_0}{A}.
\end{equation}
This relation explicitly demonstrates that the localization length is governed by the competition between the attractive interaction, encoded in the negative offset $C_0$, and the geometry--gauge renormalized stiffness $A$ generated by the helicoidal metric and the magnetic momentum shift. The semiclassical quantization of the radial motion is obtained from the Bohr--Sommerfeld condition applied to the general energy shell \cite{goldstein1950classical},
\begin{equation}
\oint p_{\xi}\,d\xi=2\pi\hbar\left(n+\frac12\right),
\end{equation}
where
\begin{equation}
p_{\xi}(\xi)=\sqrt{\mu\left[E-V_{\mathrm{eff}}(\xi;p_v)\right]} .
\end{equation}
Using the harmonic approximation,
\begin{equation}
V_{\mathrm{eff}}(\xi;p_v)\simeq C_0+A\xi^2 ,
\end{equation}
the semiclassical spectrum of the radial oscillator becomes
\begin{equation}
E_n=C_0+2\hbar\sqrt{\frac{A}{\mu}}\left(n+\frac12\right).
\end{equation}
The zero-energy sector is obtained by imposing $E_n=0$, leading to the quantization condition
\begin{equation}
\frac{g}{a}-\frac{p_v^2}{\mu}=2\hbar\sqrt{\frac{A}{\mu}}\left(n+\frac12\right),
\end{equation}
or equivalently,
\begin{equation}
p_v^2(n)=\frac{\mu g}{a}-2\hbar\sqrt{\mu A}\left(n+\frac12\right).
\end{equation}
The resulting relation defines a curvature--gauge dressed family of admissible longitudinal canonical momenta. Their existence is controlled by the combined effect of the helicoidal geometry, electromagnetic coupling, and the attractive interaction. The discreteness of the allowed values of $p_v$ emerges from the semiclassical quantization of the geometry-dependent radial motion rather than from an externally imposed confinement potential. Consequently, the helicoidal manifold acts as an intrinsic regulator of the bound-state landscape, transforming the conventional two-body problem into a nonlinear geometry--gauge controlled binding system.

\begin{figure*}[ht]
\centering
\includegraphics[scale=0.50]{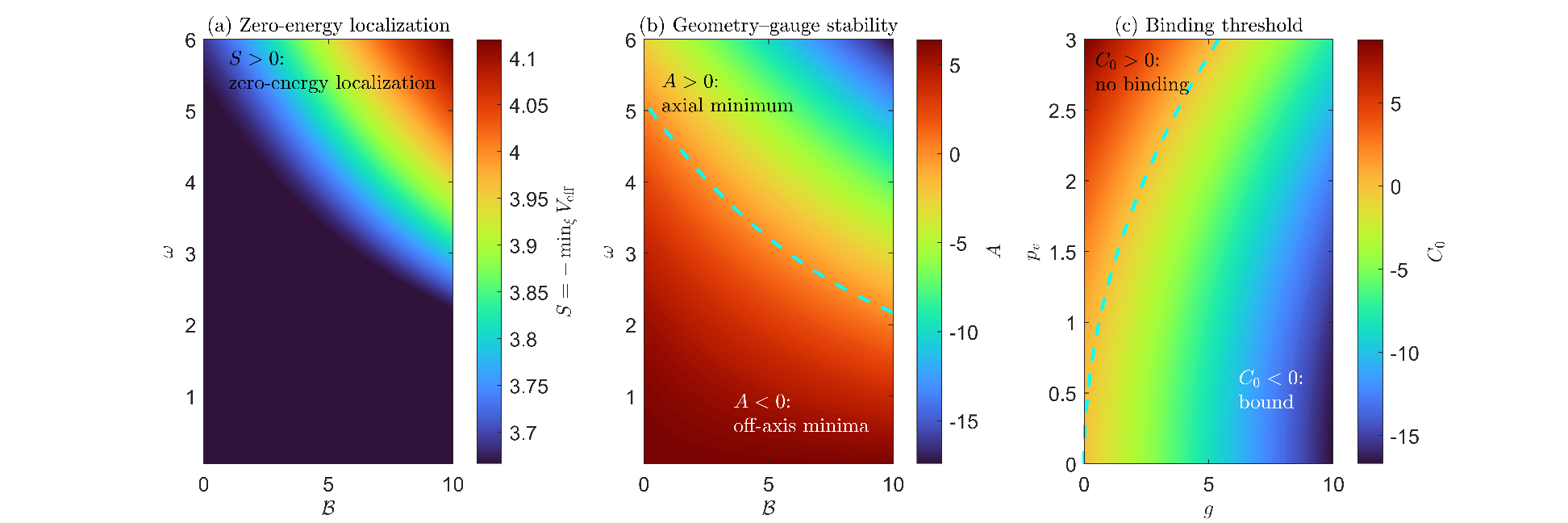}\\
\caption{\fontsize{7.6}{8.6}\selectfont  \textbf{Geometry--gauge phase structure of zero-energy localization.} Phase diagrams characterizing the emergence of zero-energy bound trajectories on the helicoidal manifold. (a) The numerical localization strength $S=-\min_{\xi}V_{\mathrm{eff}}(\xi;p_v)$ in the $(\mathcal B,\omega)$ plane, obtained from the full effective potential. Positive values of $S$ indicate the existence of a classically accessible zero-energy localization region. (b) The geometry--gauge stiffness coefficient $A$ controlling the local stability of the helicoidal axis. The line $A=0$ marks the transition between stable axial localization and symmetry-broken off-axis minima. (c) The binding threshold function
$C_{0}=p_v^2/\mu-g/a$ in the $(g,p_v)$ plane, where $C_0=0$ defines the onset of zero-energy binding. The calculations use $\mu=1$, $p_v=1$, $\kappa=0.12$, $q=1$, $a=0.6$, and $g=2.8$ for panels (a) and (b), with the corresponding parameters varied in panel (c). The solid cyan contours indicate the phase boundaries $A=0$, and $C_0=0$, respectively. The color scales represent the magnitude of the corresponding control functions.}
\label{fig:3}
\end{figure*}

\section{\mdseries{Curvature--Gauge Controlled Phase Transitions}} \label{PT}

\setlength{\parindent}{0pt}

The exact reduction developed above demonstrates that the helicoidal two-particle system is governed by the interplay of three intrinsically coupled mechanisms: curvature-induced modification of the kinetic sector, electromagnetic dressing of the conserved momentum, and localization generated by the attractive interaction. Unlike conventional confinement mechanisms in which localization is imposed through an external scalar potential, here the effective phase-space landscape emerges from the combined action of the embedding geometry and the projected gauge field. The helicoidal twist therefore acts as an intrinsic control parameter that continuously reshapes the balance between kinetic delocalization and interaction-driven localization. To quantify the competing effects, we introduce the dimensionless parameters
\begin{equation}
\lambda_{\Omega}\equiv \Omega a,\qquad \lambda_{\alpha}\equiv\frac{\alpha a^{2}}{p_v}, \qquad \lambda_g\equiv \frac{\mu g}{ap_v^{2}},
\end{equation}
where $\lambda_{\Omega}$ measures the strength of the geometric deformation on the scale of the interaction range, $\lambda_{\alpha}$ characterizes the relative importance of the gauge-induced momentum shift, and $\lambda_g$ quantifies the attractive binding strength compared with the conserved longitudinal kinetic energy. The parameter $\lambda_{\alpha}$ is defined for $p_v\neq0$ and should be interpreted as a local measure of gauge dressing. The curvature contribution modifies the effective inverse metric associated with the Killing direction according to \(g_{\mathrm{eff}}^{vv}(\xi)=\frac{1}{1+\Omega^{2}\xi^{2}}\), while the electromagnetic coupling appears through the shifted canonical momentum, \(\Pi(\xi)=p_v-\alpha\xi^{2}\). Hence, the helicoidal deformation does not merely provide a perturbative correction to the particle motion; rather, it reconstructs the effective kinetic geometry in which the reduced dynamics evolves. The existence of a zero-energy localized configuration is determined by the condition
\begin{equation}
V_{\mathrm{eff}}(\xi;p_v)\leq0 .
\end{equation}
The exact boundary separating localized and delocalized trajectories is obtained from the condition that the global minimum of the effective potential reaches the zero-energy manifold,
\begin{equation}
\min_{\xi}V_{\mathrm{eff}}(\xi;p_v)=0 .
\end{equation}
In the symmetric axial phase, where the minimum is located at $\xi=0$, this criterion reduces to
\begin{equation}
V_{\mathrm{eff}}(0;p_v)=0 ,
\end{equation}
yielding the critical longitudinal momentum \(p_v^{\,c}=\sqrt{\frac{\mu g}{a}}\). Therefore, within the axial localization regime,
\begin{equation}
|p_v|<p_v^{\,c},
\end{equation}
the attractive interaction is sufficient to compensate the longitudinal kinetic contribution. Conversely,
\begin{equation}
|p_v|>p_v^{\,c},
\end{equation}
the zero-energy bound region disappears from the vicinity of the helicoidal axis. The resulting localization threshold is thus not determined solely by the interaction strength, but emerges from the combined influence of conserved momentum, geometric deformation, and gauge coupling. The characteristic spatial extension of the localized configuration is obtained from the harmonic expansion around the stable minimum,
\begin{equation}
V_{\mathrm{eff}}(\xi)\simeq C_0+A\xi^2 ,
\end{equation}
which gives the localization length \cite{Sakurai,Gottfried}
\begin{equation}
\ell_{\mathrm{loc}}=\sqrt{-\frac{C_0}{A}},
\end{equation}
where
\begin{equation}
C_0=\frac{p_v^2}{\mu}-\frac{g}{a},
\end{equation}
and
\begin{equation}
A=\kappa-\frac{1}{\mu}\left(2\alpha p_v+\Omega^2p_v^2\right)+\frac{g}{2a^3}.
\end{equation}
This length scale characterizes the confinement radius resulting from the competition between attractive localization and geometry--gauge induced modification of the effective stiffness. In the flat-space limit, \(\Omega\rightarrow0\), the helicoidal contribution disappears and simultaneously \(\alpha=\frac{q\mathcal B\omega}{8}\rightarrow0\), so that the localization length reduces to the interaction-controlled value. For finite twist density, the additional curvature-dependent contribution \(-\Omega^{2}p_{v}^{2}/\mu\) modifies the local restoring force through the kinetic metric rather than through an independent scalar potential. The twist density therefore provides a purely geometric means of tuning the spatial extent of the bound configuration. The magnetic field introduces a second control mechanism through \(\alpha=\frac{q\mathcal B\omega}{8}\). Because the magnetic contribution enters through \(\Pi(\xi)=p_v-\alpha\xi^2\), its effect depends on the relative sign of the conserved momentum and the gauge-induced momentum shift. The mixed contribution \(-2\alpha p_v/\mu\) to the effective stiffness modifies the local stability of the axial configuration. Thus, the magnetic field provides a momentum-selective deformation of the localization landscape, producing a geometry-dependent gauge response that has no analogue in an unconstrained flat two-body system. A central consequence of the nonlinear effective potential is the emergence of a geometry-controlled bifurcation. The stability of the axial configuration is determined by the condition
\begin{equation}
A(\Omega,\alpha,p_v,g,a)=0 ,
\end{equation}
or explicitly,
\begin{equation}
\kappa-\frac{1}{\mu}\left(2\alpha p_v+\Omega^2p_v^2\right)+\frac{g}{2a^3}=0 .
\end{equation}
This relation defines a critical hypersurface in the parameter space of twist density, magnetic coupling, and conserved momentum. For \(A>0\), the helicoidal axis remains a stable localization point, \(\xi_0=0\).  When the system crosses into the regime \(A<0\) and provided that \(B>0\) the quartic term stabilizes the effective potential and two symmetry-related minima emerge,
\begin{equation}
\xi_\star=\pm\sqrt{-\frac{A}{2B}} .
\end{equation}
These two configurations correspond to opposite transverse displacements on the helicoidal surface and are connected by the exact reflection symmetry \(\xi\rightarrow-\xi \) \cite{guvendi2026phase}. The transition is therefore a geometry-induced dynamical bifurcation arising from the nonlinear restructuring of the effective potential landscape rather than from an externally applied symmetry-breaking field. The dynamical signature of this instability is obtained by expanding the motion around an equilibrium configuration,
\begin{equation}
\xi(t)=\xi_0+\eta(t),
\end{equation}
where $\eta$ denotes small fluctuations. The linearized equation becomes
\begin{equation}
\ddot{\eta}+\omega_{\mathrm{dyn}}^2\eta=0 ,
\end{equation}
with
\begin{equation}
\omega_{\mathrm{dyn}}=\sqrt{\frac{2}{\mu}\left.\frac{\partial^2V_{\mathrm{eff}}}{\partial\xi^2}\right|_{\xi=\xi_0}}.
\end{equation}
The vanishing of this frequency \(\omega_{\mathrm{dyn}}\rightarrow0 \) signals the softening of the axial mode at the bifurcation threshold and provides a direct dynamical signature of the curvature--gauge controlled localization transition. The large-distance behavior of the effective potential further confirms the existence of a confining sector. For \(|\xi|\rightarrow\infty\), one obtains
\begin{equation}
V_{\mathrm{eff}}(\xi)\simeq\left(\kappa+\frac{\alpha^2}{\mu\Omega^2}\right)\xi^2-\frac{g}{|\xi|}+\mathcal O(1),
\end{equation}
where the quadratic growth originates from the harmonic confinement and the asymptotic gauge contribution. Therefore, for finite confining parameters, the helicoidal geometry and the projected gauge coupling reinforce and reshape the confining behavior already present in the reduced Hamiltonian through the coordinate-dependent kinetic sector, while the attractive interaction provides the negative energy scale necessary for localization. These results establish a general framework in which curvature and gauge coupling jointly modulate the localization induced by the attractive interaction, as well as the stability and dynamical restructuring of constrained classical and quantum systems. In the helicoidal two-particle problem, the twist density acts as a geometric control parameter that continuously renormalizes the effective kinetic landscape, whereas the magnetic field introduces a momentum-dependent deformation of the effective potential through the gauge-shifted conserved canonical momentum. The resulting curvature--gauge critical manifold defines the boundary between distinct dynamical regimes of the reduced Hamiltonian, including axially localized configurations, symmetry-broken double-well states, and the associated softening of the lowest dynamical mode (see also \cite{guvendi2026phase}). Consequently, the localization transition is governed by the combined influence of geometry, gauge coupling, and attractive interaction within a single nonlinear Hamiltonian framework, rather than by any one of these mechanisms acting independently.

\setlength{\parindent}{0pt}

In Figure \ref{fig:3}, the phase diagrams reveal that zero-energy localization is not generated by any single confining mechanism, but emerges from the collective renormalization of the radial dynamics by geometry, gauge coupling, and attractive interaction.  The numerical localization strength in panel (a) provides the complete non-perturbative characterization of the accessible zero-energy manifold, where the boundary $S=0$ separates regions supporting real radial motion from those in which the zero-energy constraint cannot be satisfied.  The stiffness map in panel (b) identifies the local geometric mechanism underlying this behavior: the helicoidal deformation and magnetic momentum shift modify the effective quadratic confinement through the coefficient $A$, thereby controlling the curvature of the energy landscape around the symmetry axis.  Finally, panel (c) establishes the fundamental binding criterion by showing that localization requires the attractive interaction scale $g/a$ to overcome the longitudinal kinetic contribution associated with the conserved helicoidal momentum $p_v$.  Together, these results demonstrate that the helicoidal manifold acts as a dynamical regulator of two-body binding, where curvature-induced metric corrections and electromagnetic backreaction reshape the phase space of zero-energy states without introducing an external trapping potential.

\section{\mdseries{Quantum States and Curvature--Induced Spectral Reconstruction}}\label{QS}

\setlength{\parindent}{0pt}

\begin{figure*}[ht]
\centering
\includegraphics[scale=0.50]{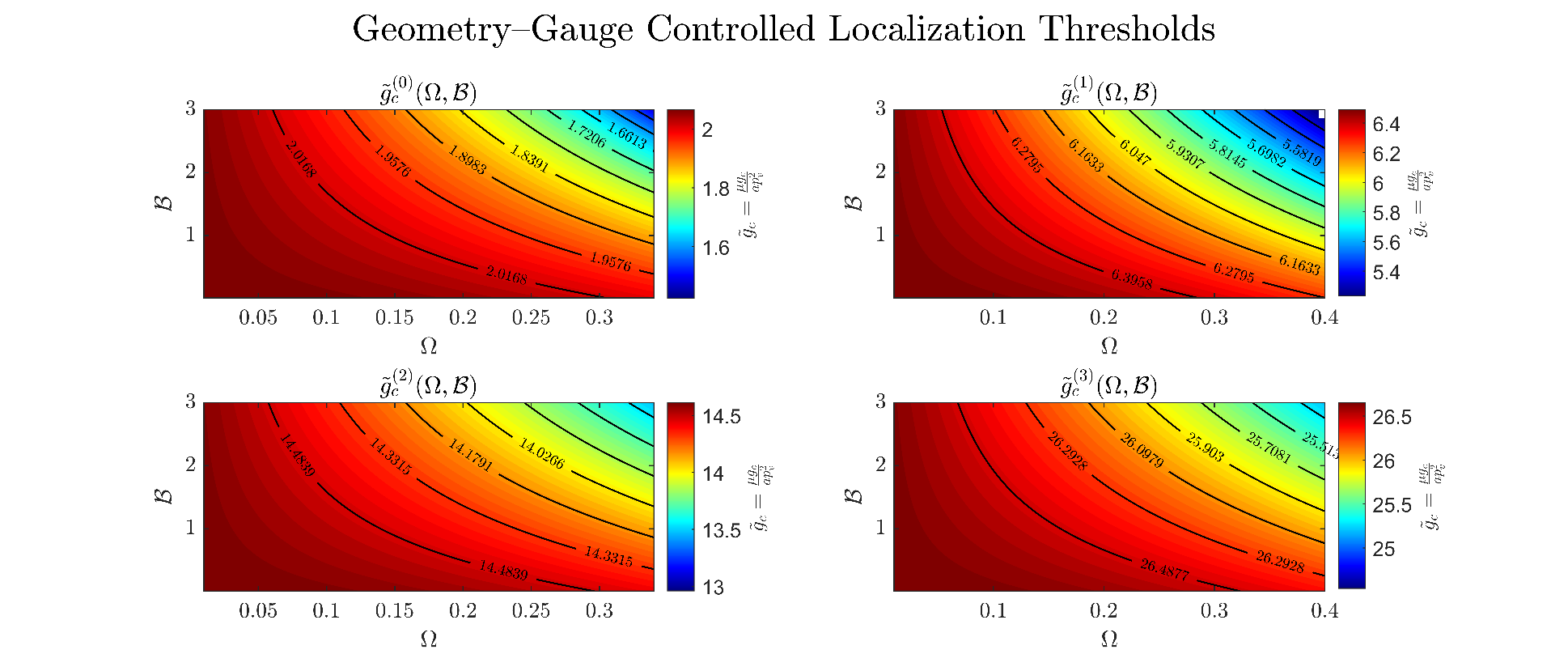}\\
\caption{\fontsize{7.6}{8.6}\selectfont \textbf{Geometry--gauge controlled quantum localization thresholds.} The contour maps display the dimensionless critical coupling
\(\tilde g_c(n)=\frac{\mu g_c(n)}{a p_v^2}\) required to support the $n$th localized quantum state below the zero-energy threshold, shown as a function of the helicoidal deformation parameter $\Omega$ and the dimensionless magnetic field $\mathcal{B}$. The critical coupling is obtained from the semiclassical threshold condition
\(g_c(n)=a\left[\frac{p_v^2}{\mu}+2\hbar\sqrt{\frac{A(g_c)}{\mu}}\left(n+\frac12\right)\right]\), where the geometry--gauge renormalized stiffness is \(A(g_c)=\kappa-\frac{1}{\mu}\left(2\alpha p_v+\Omega^2p_v^2\right)+\frac{g_c}{2a^3}\) with \(\alpha=\frac{q\mathcal{B}\Omega}{4}\). Panels correspond respectively to the quantum states
$n=0$, $n=1$, $n=2$, and $n=3$. The numerical calculations are performed using the dimensionless parameters \(\hbar=\mu=a=p_v=q=1\) and \(\kappa=0.12\). The color scale represents the minimum attractive interaction strength required for the formation of the corresponding geometry--gauge localized quantum state.}
\label{fig:4}
\end{figure*}

The classical reduction developed above provides an exact effective Hamiltonian for the relative motion of the two-body system, where the helicoidal geometry and the electromagnetic field are incorporated through the coordinate-dependent kinetic sector and the gauge-shifted conserved momentum. The quantum problem is constructed directly from this reduced phase-space description. Since the cyclic longitudinal coordinate has already been eliminated through its conserved canonical momentum $p_v$, the remaining dynamical degree of freedom is the relative helicoidal coordinate $\xi$. Therefore, the appropriate canonical quantization procedure is obtained by applying the quantum correspondence rule only to the remaining momentum variable \cite{guvendi2026charged}
\begin{equation}
p_\xi^2\rightarrow-\hbar^2\frac{d^2}{d\xi^2},
\end{equation}
while all geometry-induced contributions generated at the classical reduction stage are retained unchanged. This procedure preserves the dynamical content of the reduced Hamiltonian and avoids additional metric-dependent operator ordering ambiguities within the reduced one-dimensional formulation. Starting from the classical Hamiltonian,
\begin{equation}
\mathcal H=\frac{p_\xi^2}{\mu}+\kappa\xi^2+\frac{1}{\mu}\frac{(p_v-\alpha\xi^2)^2}{1+\Omega^2\xi^2}-
\frac{g}{\sqrt{\xi^2+a^2}},
\end{equation}
the corresponding stationary quantum equation becomes \(\hat{\mathcal H}\psi(\xi)=E\psi(\xi)\),
with
\begin{equation}
\hat{\mathcal H}=-\frac{\hbar^2}{\mu}\frac{d^2}{d\xi^2}+V_{\mathrm{eff}}(\xi;p_v),
\end{equation}
where the effective quantum landscape is
\begin{equation}
V_{\mathrm{eff}}(\xi;p_v)=\kappa\xi^2+\frac{1}{\mu}\frac{(p_v-\alpha\xi^2)^2}{1+\Omega^2\xi^2}-
\frac{g}{\sqrt{\xi^2+a^2}} .
\end{equation}
The resulting eigenvalue problem is therefore not a conventional particle in an externally prescribed potential. Instead, the effective potential is generated by the reduction of the curved two-body dynamics itself. The helicoidal geometry enters through the inverse metric factor \((1+\Omega^2\xi^2)^{-1}\), which modifies the longitudinal kinetic contribution, whereas the magnetic
field enters through the nonlinear momentum shift \(\Pi(\xi)=p_v-\alpha\xi^2 \). Consequently, the curvature and electromagnetic coupling modify the quantum localization landscape through the geometry-dependent kinetic structure and the gauge-shifted conserved momentum. To obtain the low-energy spectrum analytically, we expand the effective potential around the symmetric helicoidal axis. The required expansions are
\begin{equation}
\frac{1}{1+\Omega^2\xi^2}=1-\Omega^2\xi^2+\Omega^4\xi^4+\mathcal O(\xi^6),
\end{equation}
and
\begin{equation}
\frac{1}{\sqrt{\xi^2+a^2}}=\frac1a-\frac{\xi^2}{2a^3}+\frac{3\xi^4}{8a^5}+\mathcal O(\xi^6).
\end{equation}
Using these relations, the quantum effective potential assumes the normal form
\begin{equation}
V_{\mathrm{eff}}(\xi;p_v)=C_0+A\xi^2+B\xi^4+\mathcal O(\xi^6),
\end{equation}
where the constant energy offset is
\begin{equation}
C_0=\frac{p_v^2}{\mu}-\frac{g}{a},
\end{equation}
the curvature--gauge renormalized harmonic stiffness is
\begin{equation}
A=\kappa-\frac{1}{\mu}(2\alpha p_v+\Omega^2p_v^2)+\frac{g}{2a^3}.\label{eq:A}
\end{equation}
and the leading nonlinear correction is
\begin{equation}
B=\frac{1}{\mu}\left(\alpha^2+2\alpha p_v\Omega^2+\Omega^4p_v^2\right)-\frac{3g}{8a^5}.
\end{equation}
These coefficients provide a direct decomposition of the microscopic mechanisms controlling the quantum spectrum. The parameter $C_0$ determines the energetic position of the zero-energy threshold, while $A$ controls the local stability and quantum confinement length. Importantly, the contribution \(-\frac{1}{\mu}
(2\alpha p_v+\Omega^2p_v^2)\) shows that the helicoidal twist and magnetic coupling modify the oscillator stiffness through the kinetic sector rather than through an additional scalar potential. When the characteristic quantum localization length is sufficiently small such that the quartic contribution remains perturbative, the Hamiltonian reduces to
\begin{equation}
\hat{\mathcal H}_0=-\frac{\hbar^2}{\mu}\frac{d^2}{d\xi^2}+C_0+A\xi^2 .
\end{equation}
Because the kinetic energy has the unconventional normalization $p_\xi^2/\mu$, the corresponding oscillator frequency is
\begin{equation}
\omega_q=2\sqrt{\frac{A}{\mu}},
\end{equation}
and the characteristic quantum localization length is
\begin{equation}
\ell_q=\left(\frac{\hbar^2}{\mu A}\right)^{1/4}.
\end{equation}
The eigenstates are therefore
\begin{equation}
\psi_n(\xi)=
N_n
\exp\left(-\frac{\xi^2}{2\ell_q^2}\right)
H_n\left(\frac{\xi}{\ell_q}\right),
\qquad
N_n=
\frac{1}{\sqrt{2^n n!\sqrt{\pi}\ell_q}} .
\end{equation}
with energy levels
\begin{equation}
E_n=C_0+2\hbar\sqrt{\frac{A}{\mu}}\left(n+\frac12\right).
\end{equation}
The existence of localized quantum states below the zero-energy reference is determined by the condition \(E_n<0\), which gives
\begin{equation}
\frac{g}{a}-\frac{p_v^2}{\mu}>2\hbar\sqrt{\frac{A}{\mu}}\left(n+\frac12\right).
\end{equation}
This relation reveals that localization is controlled by a competition between the attractive interaction, the conserved longitudinal kinetic contribution, and the geometry--gauge modification of the oscillator stiffness. The critical interaction strength required to support the $n$th quantum state is therefore
\begin{equation}
g_c(n)=a\left[\frac{p_v^2}{\mu}+2\hbar\sqrt{\frac{A(g_c)}{\mu}}\left(n+\frac12\right)\right].
\end{equation}
For \(g>g_c(n)\), the corresponding quantum state becomes energetically allowed and localized around the helicoidal axis. The ground-state threshold is obtained as
\begin{equation}
g_c(0)=a\left[\frac{p_v^2}{\mu}+\hbar\sqrt{\frac{A(g_c)}{\mu}}\right].
\end{equation}
Thus, the critical coupling is not determined exclusively by the attractive interaction but is shifted by the conserved momentum and by the geometry--gauge dressing of the kinetic energy. A fundamental spectral reconstruction occurs when the quadratic confinement
coefficient vanishes, \(A=0\), which defines the critical geometry--gauge surface
\begin{equation}
\kappa-\frac{1}{\mu}(2\alpha p_v+\Omega^2p_v^2)+\frac{g}{2a^3}=0 .
\end{equation}
At this point, the harmonic localization mechanism disappears and the quartic term becomes the dominant restoring contribution,
\begin{equation}
V_{\mathrm{eff}}(\xi)\simeq C_0+B\xi^4 .
\end{equation}
The spectral dependence consequently undergoes a qualitative reconstruction from the equally spaced harmonic spectrum to the universal quartic critical scaling law,
\begin{equation}
E_n-C_0\sim\left(\frac{\hbar^4B}{\mu^2}\right)^{1/3}\left(n+\gamma\right)^{4/3},
\end{equation}
where $\gamma$ denotes the universal spectral offset of the one-dimensional quartic oscillator \cite{Sakurai,Gottfried,bender1969anharmonic,bender1973anharmonic}. The appearance of the non-integer exponent $4/3$ together with the universal offset $\gamma$ establishes that the low-energy excitations no longer belong to the harmonic universality class but instead are governed by the critical quartic fixed point generated by the complete suppression of the quadratic restoring force \cite{bender1969anharmonic,bender1973anharmonic}. Consequently, the vanishing of the geometry--gauge renormalized stiffness $A$ produces a genuine spectral criticality, in which the excitation hierarchy, localization length, and density of states are simultaneously reconstructed through the nonlinear quartic confinement. This quantum critical reconstruction constitutes the exact spectral counterpart of the classical geometry-induced bifurcation derived from the reduced Hamiltonian, thereby establishing a direct correspondence between the topological restructuring of the classical phase space and the universality class of the quantum spectrum. The helicoidal twist therefore assumes the role of an intrinsic geometric control parameter that continuously engineers localization, critical coupling, and excitation spacing through the kinetic sector alone, without introducing any external confining potential. The internal consistency of the formulation is confirmed by the flat-space limit. For
\(\Omega\rightarrow0\) and \(\alpha\rightarrow0\), the geometric and gauge dressing disappear identically, and the quantum Hamiltonian reduces to
\begin{equation}
\mathcal H_{\mathrm{flat}}=\frac{p_\xi^2}{\mu}+\kappa\xi^2+\frac{p_v^2}{\mu}-\frac{g}{\sqrt{\xi^2+a^2}},
\end{equation}
which recovers the conventional relative-coordinate two-body problem in Euclidean space. The quartic critical spectrum therefore has no flat-space analogue and emerges exclusively from the geometry--gauge reconstruction of the reduced kinetic sector. Consequently, the spectral reorganization reported here is an intrinsic manifestation of helicoidal geometry rather than a consequence of externally imposed confinement, demonstrating that engineered curved manifolds provide a fundamental mechanism for controlling quantum localization, criticality, and bound-state formation through geometry itself.

\setlength{\parindent}{0pt}

In Figure \ref{fig:4}, the geometry--gauge controlled quantum localization diagrams demonstrate that the formation of localized states is governed by the renormalization of the effective kinetic sector induced by the helicoidal geometry and electromagnetic gauge coupling, together with the attractive interaction strength. The helicoidal twist and magnetic field modify the geometry--gauge dressed oscillator stiffness \(A(g_c)=\kappa-\frac{1}{\mu}\left(2\alpha p_v+\Omega^2p_v^2\right)+\frac{g_c}{2a^3}\), thereby reconstructing the critical interaction threshold
\(\tilde g_c(n)=\mu g_c(n)/(a p_v^2)\) required for the formation of the $n$th localized quantum state. Panels correspond respectively to the states \(n=0,1,2,3\), showing the progressive increase of the localization threshold associated with higher quantum excitations. Regions of reduced \(\tilde g_c(n)\) indicate enhanced geometry--gauge assisted localization,
whereas larger values correspond to the suppression of bound-state formation due to the increased excitation energy. The resulting critical manifolds provide a direct signature of geometry-induced spectral reconstruction, demonstrating that the helicoidal structure acts as an intrinsic quantum regulator of localization by controlling the effective kinetic landscape without requiring an external confining potential.

\section{\mdseries{Conclusion}}\label{conc}

\setlength{\parindent}{0pt}

In this work, we have developed an exact curvature--gauge formulation for a two-body system constrained to a helicoidal Riemannian manifold and demonstrated that geometric deformation can serve as an intrinsic control parameter for classical and quantum localization. The central outcome of this study is that the embedding geometry modifies the dynamical structure itself rather than acting as a perturbative correction to a flat-space system. Through the exact reduction leading to the Hamiltonian in equation \eqref{eq:hamiltonian}), the helicoidal metric and the projected electromagnetic field are incorporated into a unified phase-space description in which the effective kinetic sector and the conserved momentum structure are simultaneously reconstructed. The physical origin of this behavior is encoded in two geometric quantities: the coordinate-dependent metric contribution introduced by the helicoidal embedding in equation \eqref{eq:metric} and the gauge-induced modification of the conserved longitudinal momentum. These terms transform the conventional two-body dynamics into a non-Euclidean Hamiltonian system with a position-dependent inertial response and a geometry-dependent gauge sector. Consequently, the localization properties are not determined solely by the attractive interaction, but emerge from the modified phase-space structure generated by the curved manifold and the projected gauge field.

\setlength{\parindent}{0pt}

The classical analysis reveals a qualitative restructuring of the accessible dynamical domain. The turning-point analysis presented in Figure~\ref{fig:1} demonstrates that the curvature and gauge contributions can modify the topology of the allowed phase-space regions, producing transitions between different bounded-motion regimes. In particular, the appearance of four physical turning points confirms the formation of additional dynamical barriers generated by the geometry-dependent effective potential. This result shows that helicoidal deformation provides a direct mechanism for controlling classical trajectories through modifications of the kinetic sector and the conserved momentum landscape. The same framework leads to a geometry-controlled localization mechanism at the zero-energy threshold. By introducing the finite-range attractive interaction, the effective potential given in Sec.~\ref{ZET} provides a direct criterion for the existence of localized configurations. The results of Figure~\ref{fig:2} show that the localization region can be continuously shifted through the twist density, magnetic field, and interaction strength. Importantly, the confining behavior does not originate from an additional external trapping potential; instead, it results from the modification of the effective kinetic energy and gauge structure induced by the curved manifold.

\setlength{\parindent}{0pt}

The stability analysis further identifies a critical reorganization of the localization landscape. The stiffness coefficient introduced in equation~\eqref{eq:A} determines the stability of the axial configuration and defines the critical condition separating single-well and symmetry-broken regimes. When this coefficient changes sign, the system undergoes a geometry-induced bifurcation in which the stable configuration moves away from the helicoidal axis while maintaining the reflection symmetry of the underlying geometry. The phase diagrams in Figure~\ref{fig:3} establish the corresponding critical boundaries and demonstrate that the localization transition is governed by quantitative changes in the geometry--gauge dressed effective potential.

\setlength{\parindent}{0pt}

The quantum analysis confirms that the geometric control mechanism persists beyond classical dynamics. The quantized reduced Hamiltonian developed in Sec.~\ref{QS} shows that the helicoidal deformation modifies the spectral structure through the same effective kinetic reconstruction that controls the classical trajectories. The critical interaction strengths required for different localized quantum states, presented in Figure~\ref{fig:4}, reveal that the formation of bound states is governed by a geometry-dependent spectral threshold rather than by the attractive interaction alone. Higher excitation levels acquire distinct localization conditions, demonstrating that the helicoidal geometry provides a direct means of organizing the quantum spectrum.

\setlength{\parindent}{0pt}

A particularly significant feature is the critical spectral reconstruction associated with the vanishing of the quadratic confinement term. At this point, the system leaves the harmonic localization regime and enters a quartic critical regime, establishing a direct connection between the classical soft-mode transition and the quantum spectral response. This result demonstrates that
geometric deformation can modify not only localization length scales and binding thresholds, but also the functional form of the low-energy excitation spectrum.

\setlength{\parindent}{0pt}

The results obtained here establish a general framework in which curvature and gauge structure become controllable ingredients of dynamical and quantum state engineering. Although the helicoidal manifold provides the specific realization considered in this work, the mechanism relies on general geometric principles: a nontrivial metric modifies the kinetic sector, while gauge projection reshapes the conserved quantities governing the reduced dynamics. This framework is therefore applicable to a broad class of curved electronic, photonic, and synthetic quantum systems where geometric deformation and effective gauge fields can be designed. The present results identify engineered geometry as a viable route for controlling localization, phase-space organization, and spectral properties without relying exclusively on conventional external confinement mechanisms.

\section*{\small Funding}

No funding was received for this study.

\section*{\small Availability of Data and Materials}

No data were generated or analyzed in this theoretical study.

\section*{\small Competing Interests}

The authors declare no competing interests.

\section*{\small Authors' Contributions}

\textbf{Abdullah Guvendi}: Conceptualization, Methodology, Formal Analysis, Investigation, Writing -- Original Draft, Writing -- Review and Editing, Visualization.\\

\textbf{Hassan Hassanabadi}: Conceptualization, Methodology, Formal Analysis, Investigation, Writing -- Original Draft, Writing -- Review and Editing, Visualization.\\

\end{document}